\documentclass[graybox]{svmult}

\usepackage{amsmath}
\usepackage{mathptmx}       
\usepackage{helvet}         
\usepackage{courier}        
\usepackage{type1cm}        
\usepackage{makeidx}         
\usepackage{graphicx}        
\usepackage{multicol}        
\usepackage[bottom]{footmisc}
\usepackage{color}

\makeindex             

\newcommand{\beq}{\begin{equation}}
\newcommand{\eeq}{\end{equation}}

\newcommand{\Eq}[1]{Eq.~(\ref{#1})}

\begin{document}
\title*{Dynamics on expanding spaces: modeling the emergence of novelties}
\author{Vittorio Loreto, Vito~D.~P.~Servedio, Steven~H.~Strogatz and Francesca Tria}
\institute{
Vittorio Loreto, 
					\at Sapienza University of Rome, Physics Department, Italy \& \\
					ISI Foundation, Torino, Italy \& \\
					SONY-CSL Paris, France\\ 
					\email{vittorio.loreto@roma1.infn.it}
\and Vito~D.~P.~Servedio, 
					\at Institute for Complex Systems (CNR-ISC), Rome, Italy \& \\
					Sapienza University of Rome, Physics Department, Italy\\
					\email{Vito.Servedio@roma1.infn.it}							
\and Steven~H.~Strogatz, 
					\at Cornell University, Ithaca, NY, USA \\
					\email{shs7@cornell.edu}												
\and Francesca Tria, 
					\at ISI Foundation, Torino, Italy \\ 
					\email{fratrig@gmail.com}					
}
%
%
\maketitle

\abstract{
Novelties are part of our daily lives. We constantly adopt new technologies, conceive new ideas, meet new people, experiment with new situations.  Occasionally, we as individuals, in a complicated cognitive and sometimes fortuitous process, come up with something that is not only new to us, but to our entire society so that what is a personal novelty can turn into an innovation at a global level. Innovations occur throughout social, biological and technological systems and, though we perceive them as a very natural ingredient of our human experience, little is known about the processes determining their emergence. Still the statistical occurrence of innovations shows striking regularities that represent a starting point to get a deeper insight in the whole phenomenology. This paper represents a small step in that direction, focusing on reviewing the scientific attempts to effectively model the emergence of the new and its regularities, with an emphasis on more recent contributions: from the plain Simon's model tracing back to the 1950s, to the newest model of Polya's urn with triggering of one novelty by another. What seems to be key in the successful modelling schemes proposed so far is the idea of looking at evolution as a path in a complex space, physical, conceptual, biological, technological, whose structure and topology get continuously reshaped and expanded by the occurrence of the new. Mathematically it is very interesting to look at the consequences of the interplay between the ``actual'' and the ``possible'' and this is the aim of this short review. 
}

\section{Introduction}
\label{intro}
Historically the notion of the new has always offered challenges to humankind. What is new often defies the natural tendency of humans to predict and control future events. Still, most of the decisions we take are based on our expectations about the future. The word {\it new} itself assumes many different meanings (see for instance~\cite{North_2013}). We experience novelties very often in our daily lives.  We meet new people, adopt new words, listen to new songs, watch a new movie, use a new technology. Something can be new only for us or a few other people, or something can be brand new and change a paradigm or the habits of a whole population. This is a very significant phenomenon often referred to as innovation, a fundamental factor in the evolution of biological systems, human society, and technology. From this perspective a thorough investigation and a deep understanding of the underlying mechanisms through which novelties and innovations emerge, diffuse, compete and stabilize is key to progress in all 
sectors of human activities. 

Novelties and innovations share an important feature: they can be viewed as first time occurrences of something at the individual or collective level, respectively. Though a lot is known about the way novelties and innovations can possibly emerge in specific sectors, the general picture remains poorly understood theoretically and undocumented empirically. Most of the knowledge in this area is highly scattered among either highly specialized and applied environments or academic, abstract and sometimes anecdotal publications. This paper aims at partially filling this gap by focusing on the mathematical modeling of the new and presenting a fairly incomplete review of how this problem has been tackled along with the ability of the different approaches presented to explain empirical data.

\begin{figure*}[t]
\centering
\includegraphics*[width=0.95\columnwidth]{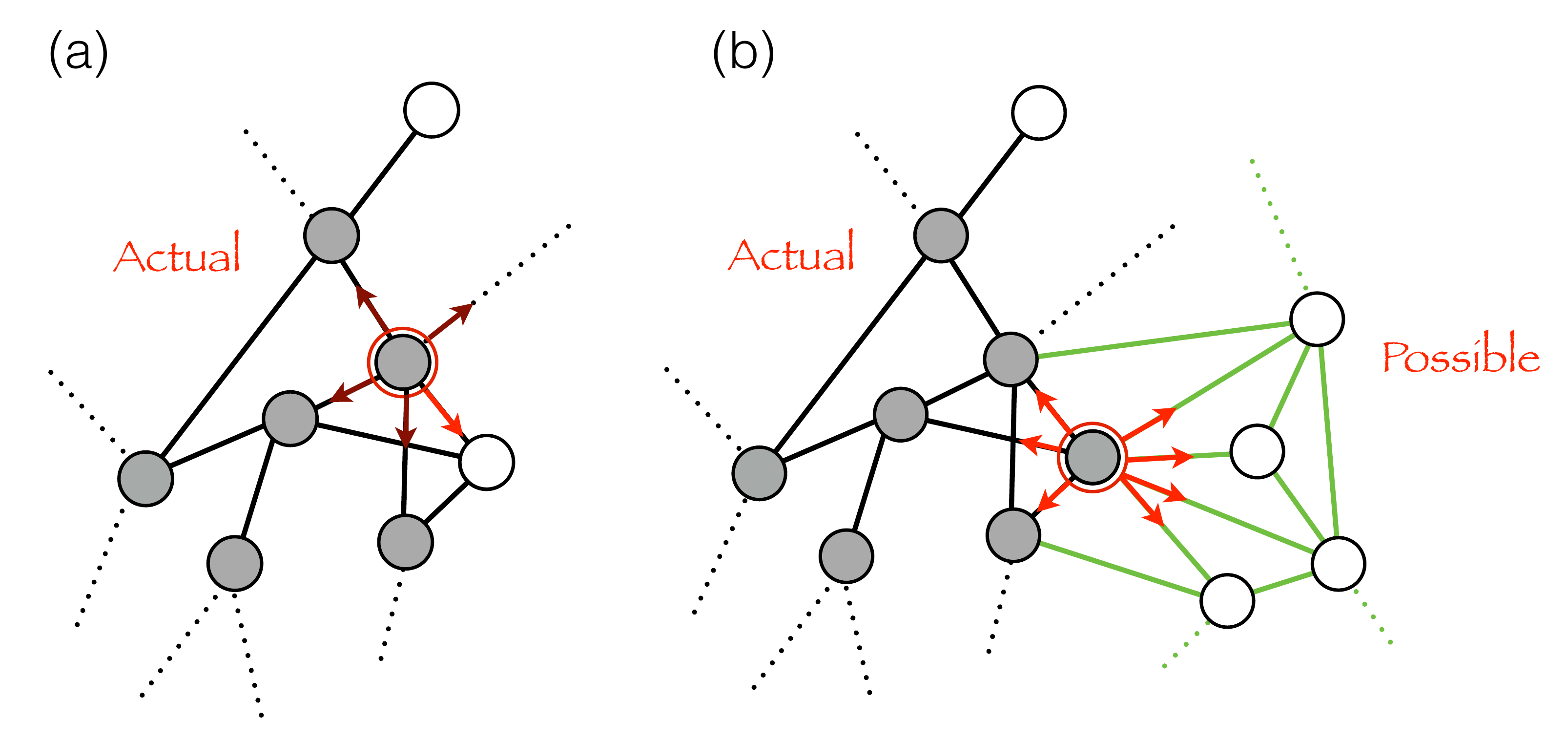}
\caption{Mathematical illustration of the adjacent possible in terms of a graph that conditionally expands from the situation depicted in (a) to that depicted in (b) whenever a walker visits a node for the first time (the white node in (a)).
\label{fig:adjacent_possible_network}
}
\end{figure*}

The reason why modeling what is new is difficult is related to a paradox that inference theories spell out very clearly. Inference is the branch of mathematics and statistics that deals with the problem of deriving logical conclusions from premises known or assumed to be true. A typical problem is that of estimating the probabilities of future events based on the observation of the past. It is interesting to report a passage from a review by Zabell on this subject~\cite{Zabell_1992}:
\begin{quote}
{ This is not the problem of observing the `impossible', that is, an event whose possibility we have considered but whose probability we judge to be $0$. Rather, the problem arises when we observe an event whose existence we did not even previously suspect; this is the so-called problem of `unanticipated knowledge'.}
\end{quote}
So, the unanticipated knowledge seems to coincide with our intuitive notion of the new. What is new is by definition out of our modeling scheme and we are left with the problem of how to incorporate its occurrence in a coherent logical scheme. Another way to look at it is through the well-known dichotomy between the actual and the possible~\cite{Jacob_1982}. A very interesting notion here is that of the {\em adjacent possible}~\cite{kauffman_1993,kauffman_2000}. Originally introduced in the framework of biology, the adjacent possible metaphor include all those things, ideas, linguistic structures, concepts, molecules, genomes, technological artifacts, etc., that are one step away from what actually exists, and hence can arise from incremental modifications and/or recombination of existing material. In Steven Johnson's words~\cite{johnson_2010_book}: 
\begin{quote}
{The strange and beautiful truth about the adjacent possible is that its boundaries grow as one explores them.} 
\end{quote}
The very definition of adjacent possible encodes the dichotomy between the {\em actual} and the {\em possible}~\cite{Jacob_1982}: the actual realization of a given phenomenon and the space of possibilities still unexplored. Fig.~\ref{fig:adjacent_possible_network} illustrates this idea with a cartoon. A walker is wandering on the nodes of a graph. The gray nodes are those already visited in the past while the white ones have  never been visited. Once the walker visits a white node for the first time, another part of the graph appears that could not even be foreseen before visiting that node.

Though the creative power of the expansion into the adjacent possible is widely appreciated at an anecdotal level, still its importance in the scientific literature~\cite{kauffman_2000,kauffman_2008,thurner_2010,Sole_2013,Felin_2014,buchanan_2014} is, in our opinion, underestimated. 

\begin{figure*}[t]
\centering
\includegraphics[width=\textwidth]{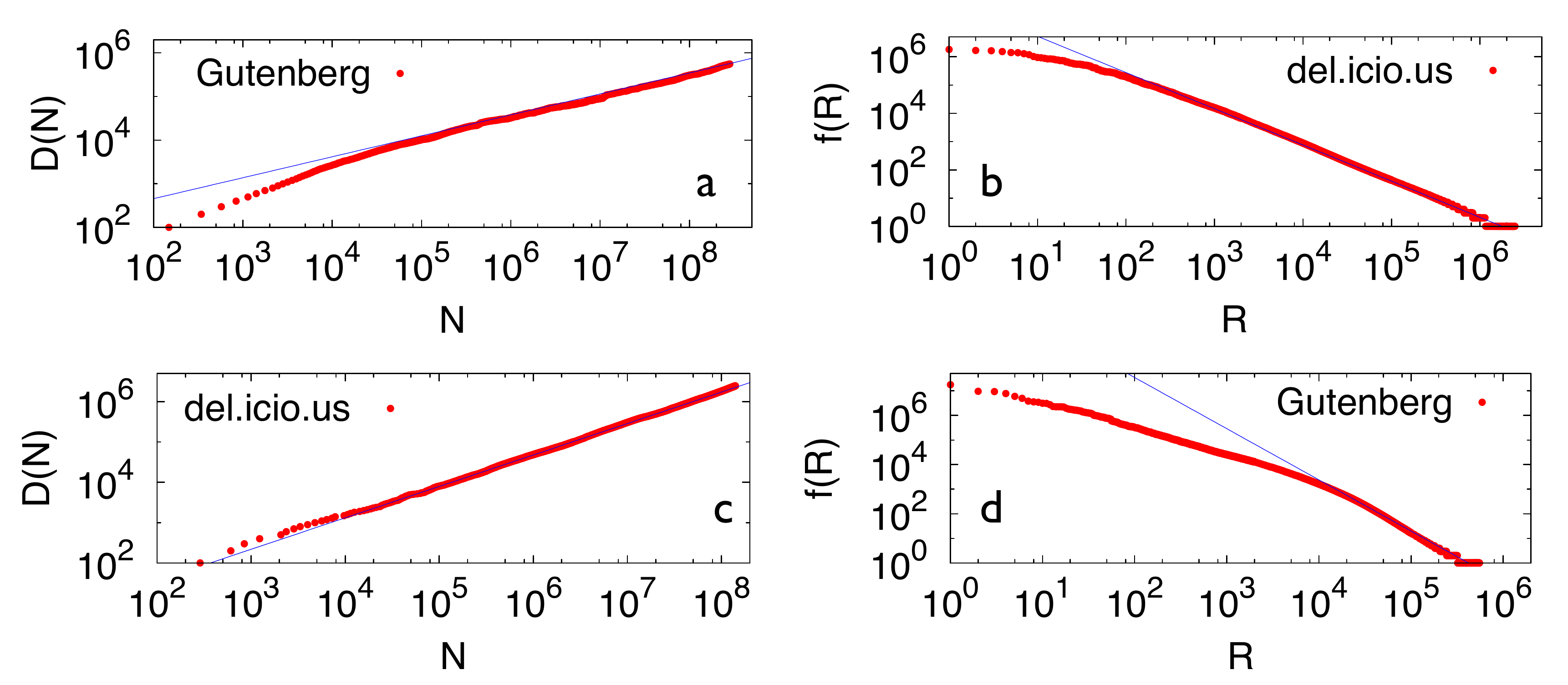}
\caption{{\bf Heaps' law (a-c) and Zipf's law (b-d) in real datasets.} Gutenberg~\cite{gutenberg} {\bf (a,b)},  del.icio.us~\cite{delicious} {\bf (c,d)} datasets.  Straight lines in the Heaps' law plots show functions of the form $f(x)=a x^{\gamma}$, with the exponent $\gamma$ equal respectively to $\gamma=0.45$ (Gutenberg) and $\gamma=0.78$ (del.icio.us). Straight lines in the Zipf's law plots show functions of the form $f(x)=a x^{-\alpha}$, where the exponent $\alpha$ is equal to $\gamma^{-1}$ for the different $\gamma$'s considered above.}
  \label{fig:heaps_zipf}
\end{figure*}

Recently, we introduced an original mathematical model of the dynamics of novelties correlated via the adjacent possible, and derived three testable, quantitative predictions from it~\cite{adjacent_possible_2014}. The model predicts the statistical laws for the rate at which novelties happen (Heaps' law) and for the frequency distribution of the explored regions of the space (Zipf's law), as well as the signatures of the correlation process by which one novelty sets the stage for another. The predictions of this model were tested on four data sets of human activity: the edit events of Wikipedia pages (Fig.~\ref{fig:heaps_zipf}(a,b)), the emergence of tags in social annotation systems (Fig.~\ref{fig:heaps_zipf}(c,d)), the sequence of words in texts, and listening to new songs in on-line music catalogs. 

By providing the first quantitative characterization of the dynamics of correlated novelties, these results provide a starting point for a deeper understanding of the adjacent possible and the different nature of triggering events (timeliness, spreading, individual vs. collective properties) that are likely to be important in the investigation of biological, linguistic, cultural, and technological evolution. As a prelude to this ambitious research program it is important to briefly summarize what is known about the attempts made so far  to model the dynamics of novelties. This is the aim of this paper.

The problem of modeling novelties is actually very old since it dates back to the work of the logician Augustus de Morgan~\cite{De_Morgan_1838}, who proposed a simple way to deal with the possibility of an unknown event. For a review of this early work as well as of more recent developments, we refer to the excellent review by Sandy Zabell~\cite{Zabell_1992}. Here, more modestly, we merely try to summarize the main steps of how investigations in this area dealt with the problem of modeling dynamical processes on expanding spaces so far. 

The outline of the paper is as follows. Section~\ref{sec:vito_simon} deals with the class of Simon-like models where the emergence of new possibilities is ruled in a probabilistic way and the rate of innovation is constant. In this class of models, though scaling is observed for the frequencies of occurrence of the different events, there is no way, except with ad-hoc solutions, to replicate empirical observations for the innovation rates. Next, in Section~\ref{sec:fra_dice} we discuss a specific example of a dynamical model on a space that shrinks, whose interest lies in the possibility to explain the emergence of scaling for the frequency distribution without resorting to a rich-gets-richer mechanism (as in the Simon-like class of models). Section~\ref{sec:fra_hoppe} describes the attempts made to introduce the emergence of innovations in Polya-like models. We mainly discuss the Hoppe-Polya model and contrast its predictions with those of similar modeling schemes. In this case we observe for the first time 
an expansion of the space of possibilities conditioned on the occurrence of a specific event. Section~\ref{sec:vittorio_polya_urns} introduces our modeling scheme for the adjacent possible. In this case the expansion of the space of possibility, i.e., the adjacent possible, is triggered in a self-consistent way by the occurrence of new events. Finally, the concluding section summarizes the main conclusions and tries to highlight interesting future directions. 

Throughout the paper, when the relevant statistical quantities are well described by power-laws, we shall indicate with $\beta$ (minus) the exponent of the frequency distribution of tokens, with $\alpha$ (minus) the frequency-rank exponent and with $\gamma$ the Heaps' exponent. By simple arguments reported in the Appendix Sec.~\ref{sec:fra_zipf_heaps}, one finds that $\alpha = (\beta-1)^{-1}$ and that $\gamma=\alpha^{-1}$ when $\alpha>1$ while $\gamma=1$ otherwise.

\section{Simon-like models}
\label{sec:vito_simon}
The observation that the frequency distribution of words in texts written in a given language follows a fat-tailed distribution has been puzzling the scientific community since the beginning of the 20th century \cite{estoup16,condon28,zipf35} and continues to be an hot topic nowadays \cite{ramon2010}.
The search for a suitable model that could reproduce the  experimental data of word frequency has also caused rather tough scientific disputes \cite{simon-mandel-dispute}.
As we will briefly show in the following, almost all models based on Simon's model are not able to generate a set of tokens whose frequency distribution is represented by a power-law with an exponent $\beta<2$ and consequently the associated Heaps' law is linear.
The only exception is the model proposed by Zanette and Montemurro \cite{zanette2005}, whereas in that case the sub-linear Heaps' exponent has to be recovered by data and inserted by hand  without a first principle explanation.

\subsection{Plain Simon's model}
\begin{figure*}[t]
\centering
\includegraphics[width=0.9\textwidth]{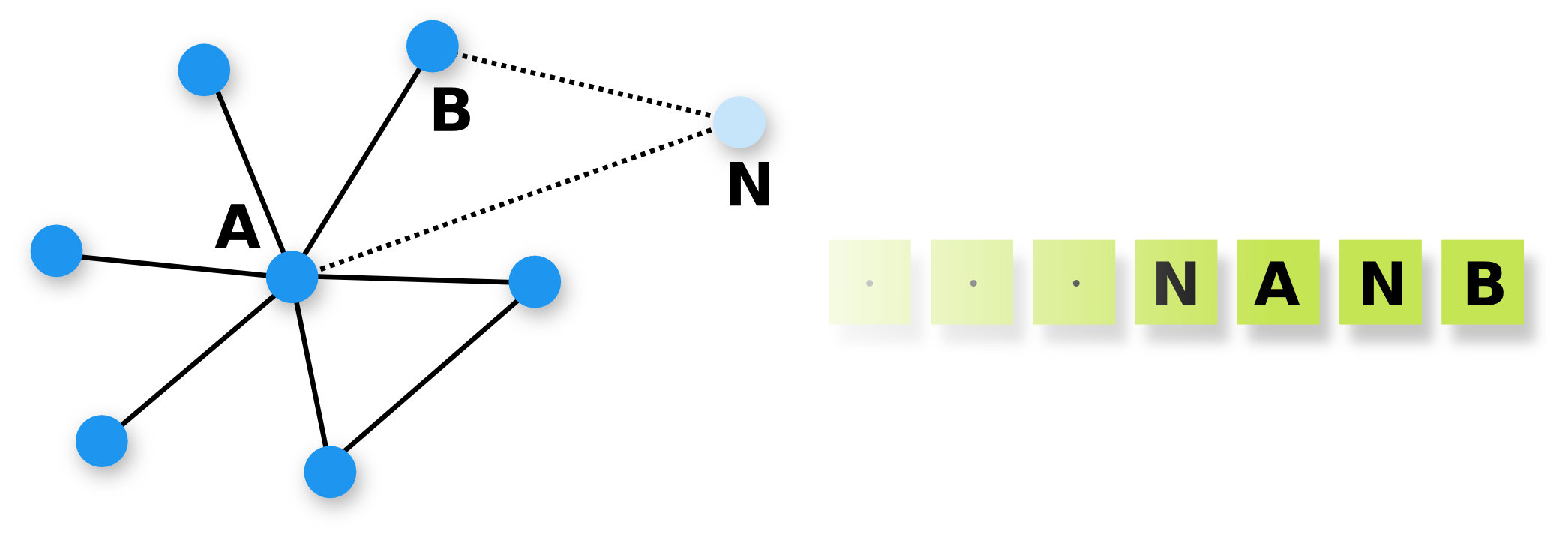}
\caption{ 
  \textbf{Equivalence between network growth models and Simon's model.}
  In a network growth model like that of Barab\'asi-Albert a new node N enters the graph and attaches to both nodes A and B. This corresponds to inserting the tokens A and B and twice N in the stream of the equivalent Simon's model in which the \emph{rich-gets-richer} mechanism happens deterministically any two time steps. Therefore, the equivalent Simon's model is obtained with the probability of extracting an old token set to $(1-p)=1/2$. The analogy is complete by identifying the number of occurrences of the tokens in Simon's stream with the connectivity of nodes in the graph. 
\label{fig:simon_graph}}
\end{figure*}

One simple model able to reproduce in part the phenomenology of texts is the stochastic model devised by Simon \cite{simon-mandel-dispute}. 
In Simon's model a stream of tokens is generated according to the following two prescriptions: 
at the beginning, i.e.,\ at time $t=1$, only one token is present in the stream; 
at a generic time $t$ a new token is added to the stream with probability $p$, while with complementary probability $(1-p)$ a randomly extracted token of the stream is chosen.
In this way the tokens that appear more frequently in the stream have a higher probability to be extracted.
This mechanism of favoring those elements that occur more frequently in the stream is called \emph{rich-gets-richer} and has become a paradigm for the generation of tokens whose frequency is distributed according to a power-law~\cite{simkin2011}. 
Because of its sequential nature, Simon's model is particularly suitable to describe phenomena of linguistics, such as the generation of texts, although there are still some linguistics aspects that cannot be reproduced.
Above all, it is evident that the rate of addition of new tokens is constant in time ($p$), thus resulting in a linear growth of the available space, i.e.,\ a linear growth of the number of different tokens $w=pt$, whereas in real texts such growth is found to be asymptotically sub-linear $w=cp^\gamma$ with $0<\gamma<1$ (Heaps' law \cite{heaps78}).   
The rich-gets-richer mechanism at the basis of Simon's model, as it often happens in science \cite{simkin2011}, has been recycled in other contexts. 
Notably, the \emph{preferential attachment} rule introduced in~\cite{barabasi99}, is effectively a disguised form of the rich-gets-richer.
In fact, in the Barab\'asi-Albert model of network generation, a new node of the graph is introduced at time $t$ by connecting it to  $m$ existing nodes chosen with probability proportional to the number of their first neighbors. 
This effectively corresponds to a deterministic Simon's process with the probability of extracting an old token set to $(1-p)=1/2$ (see Fig.~\ref{fig:simon_graph}), deterministic since at every even time step there is always a rich-gets-richer in action, while in Simon's model the entrance of an old token in the stream at any time is not certain but is conditioned to the extraction of a random number. 
The equivalent Simon's model so defined is characterized by a dictionary size increasing linearly in time as $D(t)=pt/m$.

The mathematical determination of the exponent characterizing the power-law distribution of the token occurrences in the Simon's model is not particularly difficult \cite{simon-mandel-dispute,simkin2011}.
Here, we would like to present an alternative method that involves the master equation of the process.
Although this method is not rigorous, it can be straightforwardly generalized to more difficult setups.
It is based on the master equation and the continuous approximation.
We denote with $N_{k,t}$ the number of tokens that have occurred $k$-times at time $t$ and we can write its stochastic evolution in time as 
\beq
	N_{k,t+1} =  N_{k,t} + (k-1)(1-p)\frac{N_{k-1,t}}{t} - k (1-p) \frac{N_{k,t}}{t}  + p\,\delta_{k,1},
	\label{eq:SimonMaster}
\eeq
i.e., the number of tokens occurring $k$-times at time $t+1$ equals the number of tokens occurring $k$-times at time $t$ plus the contribution we would have by extracting a token that has already occurred $(k-1)$-times in the sequence (and this happens with probability $(1-p)$ times the fraction of tokens occurring $(k-1)$-times at time $t$), minus the contribution of extracting a token already occurring $k$-times (which will contribute to increase the number of tokens occurring $(k+1)$-times), plus the specific contribution of creating a brand new token.
At time $t$ there are $t$ tokens in the stream by definition of the model, so that the probability of choosing a token occurring $k$-times is in fact $kN_{k,t}/t$.
The continuous approximation is defined as 
\beq
	\left\{
		\begin{array}{l}
			  N_{k,t+1} - N_{k,t} \approx \frac{\partial N_k}{\partial t}\\
			  kN_{k,t} - (k-1)N_{k-1,t} \approx \frac{\partial (kN_k)}{\partial k}
		\end{array}	\right.
		\label{eq:contapprox}
\eeq
and the master equation (\ref{eq:SimonMaster}) becomes
 \beq
		\frac{\partial N_k}{\partial t} = -(1-p) \frac{1}{t} \frac{\partial (kN_k)}{\partial k}.
		\label{eq:me_cont}
\eeq
It is verified numerically that  the expression $N_{k,t}/t$ tends asymptotically  to a stationary distribution $q_k$.
Therefore, by posing $N_{k,t}=tq_k$ in Eq.~\ref{eq:me_cont} and by simplifying the partial derivatives, we get
\beq
		(1-p) k \frac{dq_k}{dk} = -(2-p) q_k.
		\label{eq:me_pconst}
\eeq
By solving the previous ordinary differential equation with standard methods we obtain 
\( q_k \propto k^{-1-\frac{1}{1-p}}\), 
i.e, a power-law distribution with exponent  $\beta = 1+\frac{1}{1-p}$ corresponding to a frequency-rank exponent $\alpha=(1-p)$.
The aforementioned case of the Barab\'asi-Albert model corresponds to setting $p=1/2$, with a resulting network with an asymptotic distribution of node degree (number of first nearest neighbors) obeying a power-law with exponent $\beta=3$.

\subsection{Simon's model with time dependent sublinear invention probability}
As mentioned above, the Simon's model is in some sense satisfactory but not conclusive and two main issues can be pointed out.
First, with $p$ in the range between 0 and 1, frequency-rank exponents $\alpha$ larger than 1 cannot be recovered although lots of idioms actually display them.
Second, the dictionary, i.e., the number of different tokens (which can be interpreted as the size of the space) grows linearly in time and not sub-linearly, i.e., faster than in reality.
To correct both issues, a time dependent and decreasing probability $p_t=p_1 t^{\gamma-1}$ with $0<\gamma<1$ can be introduced in the model \cite{zanette2005}, thus assuring that the dictionary $D(t)$ would grow as $t^\gamma$, since by definition $D(t)=\int_1^t p_1 {s}^{\gamma-1} ds$. 
To see how in this case the frequency-rank distribution behaves at large $t$, we start from Eq.~(\ref{eq:me_cont}), set $p=p_t=p_1 t^{\gamma-1}$ and look for an asymptotic solution of the type $N_{k,t}=t^\gamma q_k$.
After some algebra we come to the equation
\beq
	(1-p_1 t^{\gamma-1}) k \frac{dq_k}{dk} = -(\gamma+1-p_1 t^{\gamma-1}) q_k
\eeq
that at large times becomes
\beq
	k \frac{dq_k}{dk}=-(1+\gamma) q_k.
\eeq
Its solution is then for large $t$, $q_k \propto k^{-1-\gamma}$ with a frequency-rank distribution decreasing as a power-law of exponent $1/\gamma$.
At fixed large $t$ the frequency of tokens can be written as $f_k=N_{k,t}/t\propto t^{\gamma-1} q_k$ with the time dependent factor that will be absorbed by the normalization constant.

\subsection{Simon's model with memory}
In Simon's model there is no need to introduce an explicit time ordering since the mechanism of extraction of old tokens from the stream is time independent. In fact it would be more appropriate to name the stream as ``heap'' instead.
This feature is partially justified as soon as the words of texts are considered, but when the tokens have to represent other aspects of life, things may be different.
In particular when tokens are identified with the songs listened to by one person or technological products on the market purchased by people, the age of tokens does matter in choosing them (e.g., nobody would buy an outdated  computer).

An effect of aging can be included in Simon's model in many ways. Two of them are worth of mentioning.
Dorogovtsev and Mendes \cite{mendes2000} introduced a mechanism of aging beside the preferential attachment in the Barab\'asi-Albert model of network growth. Their findings can be interpreted in the framework of Simon's model by the already mentioned equivalence of network growth processes and Simon's model with $p=1/2$ (Fig.~\ref{fig:simon_graph}). 
In their model a new node is attached to old nodes proportionally to their connectivity and their age. 
The latter is taken into account by means of a factor $\tau^{-\nu}$, with $\tau$ being the elapsed time since the token's first appearance. 
Interestingly, they find that for $\nu <1$ the connectivity of nodes still obeys a power-law distribution with exponent ranging from $-2$, obtained as $\nu \rightarrow -\infty$, and infinity as $\nu \rightarrow 1$. When $\nu >1$ the connectivity is described by an exponential distribution.
We expect the same to hold in the general case of Simon's model for any $p$, where an already occurring token can be chosen with probability $(1-p)$ proportionally to its number of occurrences and proportionally to its age to the power $-\nu$. 
We conjecture that in the general case of $p\neq 1/2$ the exponent of the frequency distribution would be bounded  above by $-2$, being exactly $\beta= -1-\frac{1}{1-p}$ in the case of $\nu=0$.
It is interesting to note that the nontrivial exponential behavior is detected whenever the aging kernel $\tau^{-\nu}$ is not an integrable function of $\tau$, i.e., when $\sum_{\tau=1}^T \tau^{-\nu}$ diverges at large $T$ so that the effective size of the sampling window is infinite.
The model of Dorogovtsev and Mendes still presents the same issues as Simon's model, i.e., the Heaps' law is linear and the exponent of the frequency distribution of tokens may not exceed the value $-2$.

\begin{figure*}[t]
\centering
\includegraphics[width=0.8\textwidth]{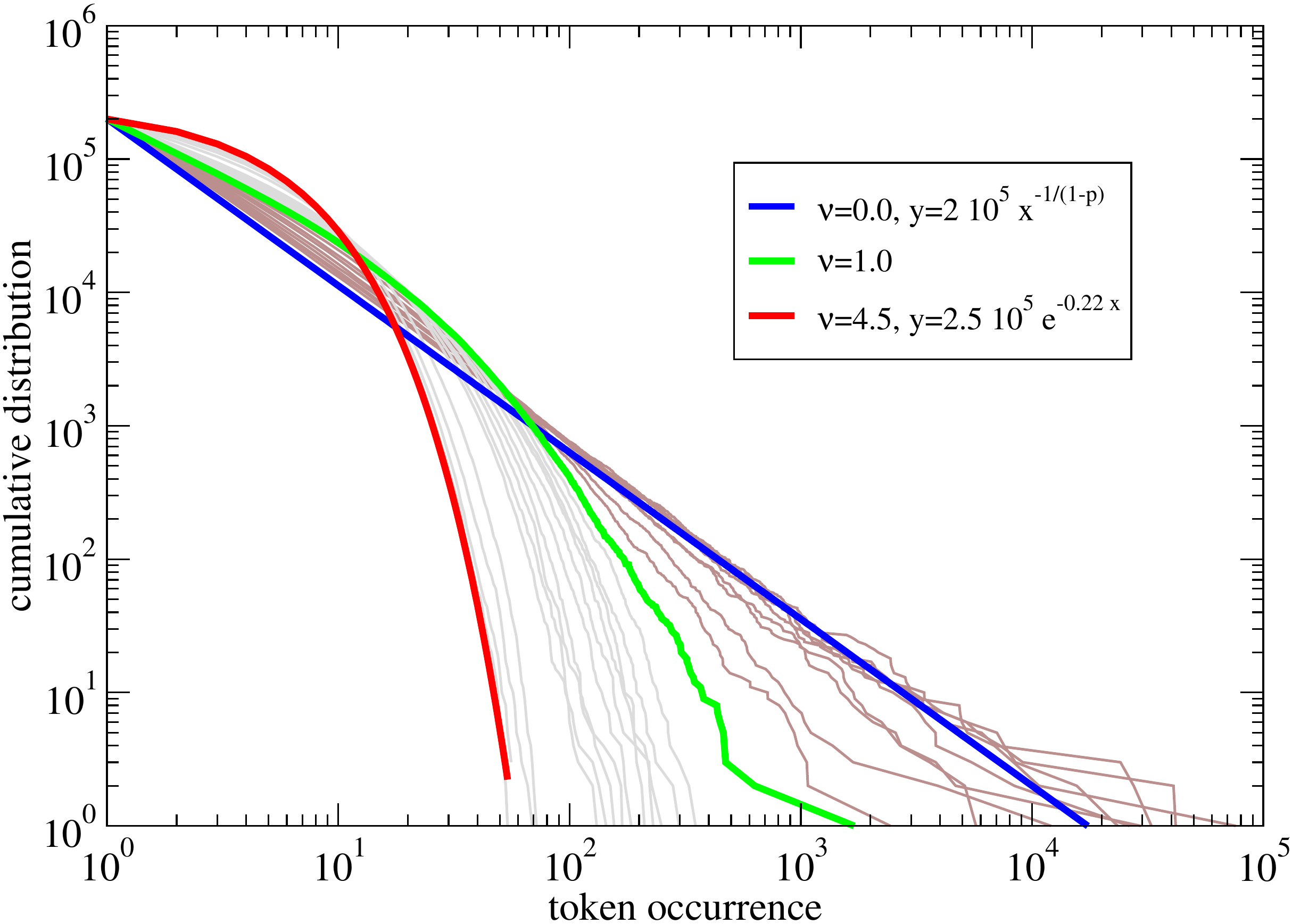} 
\caption{ 
  \textbf{Behavior of the cumulative distribution of token occurrences in the Cattuto-Loreto-Pietronero (CLP) model \cite{cattuto_pnas} as a function of the memory kernel power.}
  A stream with 1,000,000 tokens was generated with a probability of invention $p=0.2$ and a memory kernel $(1+\Delta t)^{-\nu}$ at various values of $\nu$.
  The blue line corresponds to the plain Simon's model ($\nu=0$) where the cumulative frequency distribution of tokens tends asymptotically to the power-law $f^{-1/(1-p)}$;
  The brown curves depict the results for $0<\nu<1$;
  The green curve corresponds to the stretched exponential discussed in the CLP paper ($\nu=1$);
  The gray curves are calculated for $1<\nu<4.5$;
  The red curve shows the exponential behavior $\approx e^{-0.22 f}$ found with $\nu=4.5$.
\label{fig:memory}}
\end{figure*}

A different mechanism of aging was introduced by Cattuto et al.~\cite{cattuto_pnas,cattuto_epl} to mimic the correlations and frequency of tags in social annotation systems like \emph{del.icio.us} \cite{delicious}.
In that case, the choice of the old existing token in the stream was conditional to the usual probability $(1-p)$ and to a memory kernel of the type $(r+\Delta t)^{-1}$, where $r\ge1$ is a constant and $\Delta t$ is the age of single tokens measured in time steps, i.e., tokens are no longer considered as a whole as in the Dorogovtsev-Mendes model. 
The usual Simon's model would correspond to a trivially constant kernel. 
This model results in a rather complicated token frequency distribution that resembles a power-law of exponent $-2$ and a frequency-rank distribution that is well described by a stretched exponential. 
Though in the original paper only a hyperbolic kernel was considered, it can be shown that, as in the Dorogovtsev-Mendes model, meaningful results are obtained with nonintegrable kernels. 
In fact, Fig.~\ref{fig:memory} shows that the cumulative frequency distribution of tokens displays a fat-tail behavior when kernels are of the type $(1+\Delta t)^{-\nu}$ with $\nu<1$.
In this model, the Heaps' law is again linear and the power-law frequency distribution of tokens displays an effective exponent not exceeding the value of $-2$.

{\renewcommand{\arraystretch}{1.2}
\begin{table}[t]
\begin{center}
 \begin{tabular}{|c|p{4cm}|p{4cm}|c|}
\hline
  \textbf{~~Model~~} & \centering \textbf{Characteristics} & \centering \textbf{\boldmath Zipf's exponent $\alpha$} & \textbf{Heaps' exponent}\\
\hline
\hline
Simon & Constant invention rate $p$ & \centering $1-p$ & 1\\
\hline
ZM & Variable invention rate $p=ct^{\gamma-1}$ with $0<\gamma<1$& \centering $1/\gamma$ & $\gamma$\\
\hline
DM & Aging of tokens $\tau^{-\nu}$, constant invention rate & \centering $2<\alpha<\infty$ for $-\infty<\nu<1$ & 1\\
\hline
CLP & Stream aging $(r+\Delta t)^{-1}$, constant invention rate & \centering stretched exponential with $\alpha\approx 1$& 1\\
\hline
 \end{tabular}
\caption{\textbf{Comparison of Simon-like models:}
  ZM stands for Zanette-Montemurro \cite{zanette2005}; DM for Dorogovtsev-Mendes \cite{mendes2000}; CLP for Cattuto-Loreto-Pietronero \cite{cattuto_pnas}.
\label{tab:models}}
\end{center}
\end{table}
}

\section{The sample-space reducing model}
\label{sec:fra_dice}
In the attempt to explain fat-tails distributions as observed in real
systems, an alternative method to reproduce power-law frequency
distributions without explicitly resorting to a rich-gets-richer mechanism
was proposed~\cite{Thurner_2015}.  
The model interestingly catches the idea that the space of possibilities
often locally reduces when the process goes on: for instance, when
composing a sentence the first word is almost free, while the
subsequent ones are more and more constrained.
The very simple process proposed in~\cite{Thurner_2015} works as
follows: 
(i) the process starts with an $N$-faced dice; 
(ii) at time $t$,  a $j$-faced dice resulting from the evolution of the initial $N$-faced dice is thrown, 
and let $i$ be the face value obtained; 
(iii) at time $t+1$ an $i$-faced dice is then thrown, and the process goes on until the face value 1 is extracted. 
It can be shown~\cite{Thurner_2015}
that, independently of $N$, the visiting probability for the site $i$,
defined as the probability that a particular process visits the site
$i$ before ending at 1, is:
\begin{equation}
 P_N(i)= \frac{1}{i}  \,.
\end{equation}
If we consider a  cyclic process, i.e.\ when 1 is reached the process
starts again from an $N$-faced dice, 
the visiting probability is also proportional to the occupation
probability of a site, 
and thus to the frequency rank, reproducing  an exact Zipf law $f(R)\propto
R^{-1}$. 
By relaxing the constraint of deterministically reducing the
sample space,  in~\cite{Thurner_2015}
the authors also study the case in which a probability
 $\lambda$ exists to come
 back at the $N$-faced dice at each step. In this case, the model can
 be easily studied as a superposition of the pure sample-space reducing
 process (with probability $1-\lambda$) and of a random process in which a number is drawn uniformly
 in the interval $[1, N]$ (with probability $\lambda$).  It is shown
 that one obtains in this case a generalized Zipf law with frequency-rank distribution
 $f(R)\propto R^{-\lambda}$.

\begin{figure*}[t]
\centering
\includegraphics[width=0.47\textwidth]{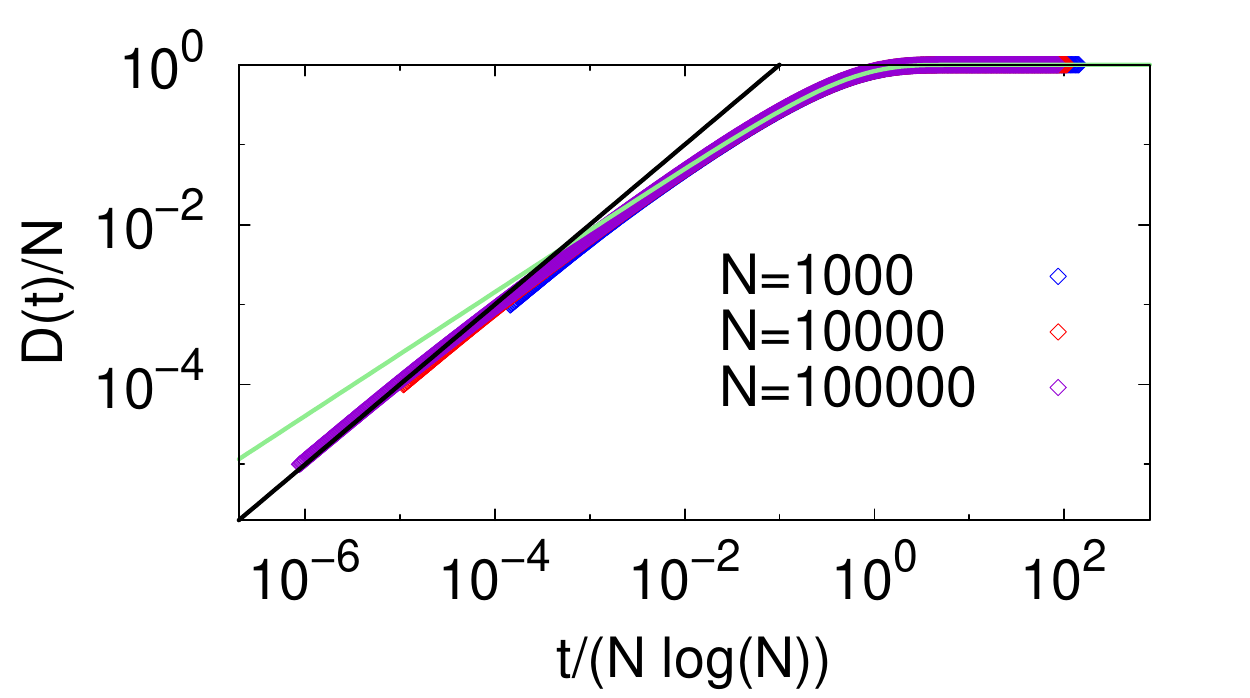}%
\includegraphics[width=0.47\textwidth]{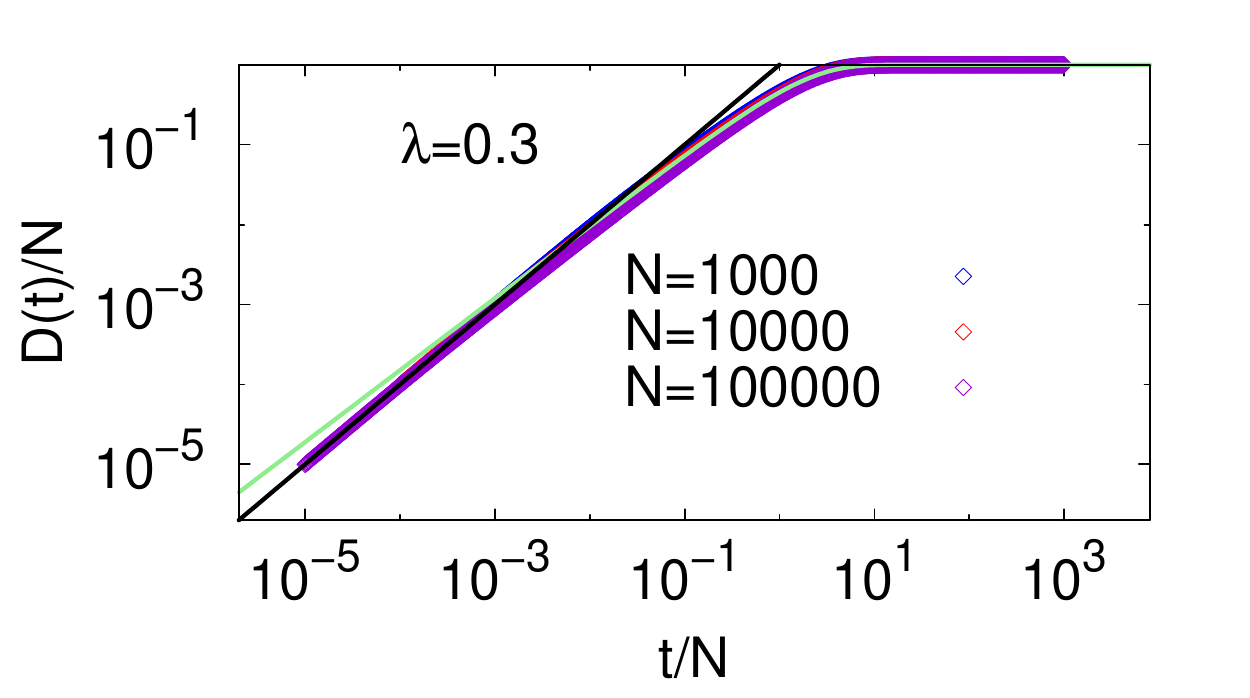}
\caption{ {\bf Heap's law in the sample-space reducing model.} Left:
  model with $\lambda=0$. Right: model with $\lambda=0.3$. In both
  cases results are shown for $N=1000,10000,100000$, and the x and y
  axes are suitably normalized in order to superimpose the different
  curves. The curves with both $\lambda=0$ and $\lambda=0.3$ are
  well fitted with a constant plus a stretched exponential function: $f(x)=(1-
  \exp(-cx^b))$, with  $ c=1.8, b=0.775$ for $\lambda=0$, and  $c=0.6,
  b=0.9$ for $\lambda=0.3$.  In both figures the straight black
line corresponds to $y\propto x$.
\label{fig:Heap_dadi}}
\end{figure*}

We consider here also the exercise of computing the average number of different values
$D(t)$ drawn until the $t$-th
step of the cyclic sample-space reducing process described above.
This obeys the equation:
\begin{equation} \label{eq:Dt}
D(t+1) -D (t) = \sum_i p_i (t) p_{i=\mathrm{new}} (t) \sim \sum_i \frac{A}{i} \left(1-
\frac{A}{i} \right)^t  \, ,
\end{equation}
where $ p_i (t) $  is the probability of extracting the
number $i$ at step $t$ and  $ p_{i=\mathrm{new}} (t) $ is the probability
that $i$ is extracted at step $t$ for the first time.  Note that the last
equivalence is not exact since we  do not consider the fixed directional
constraints imposed by the model.
 By approximating with the continuous limit, we obtain  (see
 Fig.~\ref{fig:Heap_dadi}, left panel):
\begin{equation} \label{eq:Dt_cont}
\frac{d D(t)}{dt} \sim \sum_i \frac{A}{i}  e^{-\frac{t A}{i} } \, \,\,
\Rightarrow \, \,\,  D(t)= N - \sum_i  e^{-\frac{t A}{i} }
\,
\end{equation}
where $A$ is the normalization constant:
\begin{equation}
A^{-1}= \sum_i^N \frac{1}{i} \,.
\end{equation}
Eqs.~(\ref{eq:Dt}),~(\ref{eq:Dt_cont}) can be easily generalized
when  a probability $\lambda$ to come
 back to the $N$-faced dice at each step is introduced:
\begin{equation}
\frac{d D(t)}{dt}= \sum_i (1-\lambda) \frac{A}{i} + \frac{\lambda }{N}\left(1-((1-\lambda) \frac{A}{i} + \frac{\lambda }{N}
) \right)^t \, ,
\end{equation}
resulting in (Fig.~\ref{fig:Heap_dadi}, right panel):
\begin{equation}\label{eq:Dt_dadigen}
D(t)= N -  e^{ -\frac{\lambda t}{N} }\sum_i e^{ -\frac{A t}{i}}  \,.
\end{equation}
Despite its interest, this basic version of the space
reducing model is not able to reproduce most of the interesting
phenomenology of real processes, in particular a sublinear power-law
behavior for the number of novelties introduced in the process as a
function of the number of events, and a Zipf's law with  a tail exponent
larger than one (in absolute value).

\section{Hoppe Urn Model}
\label{sec:fra_hoppe}
In 1984 Fred M. Hoppe~\cite{Hoppe_1984} introduced for the
first time the
concept of novelties in the framework of urn models~\cite{polya_1930,johnson1977urn,mahmoud_polya}, proposing what
is now called the ``Hoppe urn model''. 
The inspiration for his work
was the Ewens' sampling formula~\cite{Ewens_1972}. It describes the
 allelic partition of a random sampling of $n$ genes from an infinite
 population at equilibrium, that evolves according to a discrete time
 neutral Wright-Fisher process~\cite{fisher1930,wright1931evolution} with
 constant mutation rate $\mu$ per gene. When taking the infinite
 population limit $N \rightarrow \infty$ and $\mu \rightarrow 0$, with
 $N\mu$ constant, Ewens~\cite{Ewens_1972} showed a remarkable property
 of the sampling process: if one samples $j$ genes from the population
 at equilibrium, then the probability of sampling the $( j+1)$th gene
 with an allele already sampled is $j/( j+\theta)$, where $\theta =
 4N\mu$. 
Hoppe devised his model starting from this property. 
In the classical version of the Polya urn model~\cite{polya_1930}, balls of various colors are placed in an urn. A ball is withdrawn at random, inspected, and placed back in the urn along with a certain number of new balls of the same color, thereby increasing that color's likelihood of being drawn again in later rounds. The resulting rich-gets-richer dynamics leads to skewed distributions~\cite{simon-mandel-dispute,Yule_1925} and has been used to model the emergence of power laws and related heavy-tailed phenomena in fields ranging from genetics and epidemiology to linguistics and computer science~\cite{simkin2011,Mitzenmacher_2003,newman_2005}. 
In~\cite{Hoppe_1984,Hoppe_1987}  Hoppe  considered a Polya urn with
balls of two different qualities: one black ball of mass $\theta$, and
colored balls with mass one. The dynamical process works as
follows: 
it starts with only the black ball in the urn, then balls are
randomly chosen from the urn proportionally to their mass.
 At each time step $t$,
(i) if the black ball is extracted, a ball with a brand new color is
added to the urn together with the black ball, 
(ii) if a colored ball is extracted, it is returned
to the urn with an additional copy of it. 
It is easy to see that the probability of extracting an already existing color from the urn at each step $t +1$ is exactly $P_\mathrm{existing}(t +1) = t/(t +\theta)$, reproducing the result in~\cite{Ewens_1972}. The expected number of different colors in the urn at step $n$ can be computed explicitly and reads~\cite{Hoppe_1984,Hoppe_1987}:
\begin{equation}\label{eq:hoppe_exact}
E(k) = \frac{\theta}{\theta}+   \frac{\theta}{\theta+1}+
\frac{\theta}{\theta+2} + ... +  \frac{\theta}{\theta+t-1} \,.
\end{equation}
We will give here an alternative derivation, approximate but
straightforward, for the expected number $D(t)$ of different colors
in the urn at step $t$.

The number of different colors $D(t)$ after $t$ extractions follows the
recurrence equation:
\begin{equation}
D(t+1) = D(t)+ \frac{\theta}{\theta +t} \,,
\end{equation}
where the last term is the probability of extracting a brand
new color at time $t$. We can now take the continuous limit:
\begin{equation}
\frac{d D(t)}{  dt } = \frac{\theta}{\theta+t} ~,  ~~~ D(0)=0 \,,
\end{equation}
with solution 
\begin{equation}\label{eq:Dt_hoppe}
D(t) = \theta \ln {(\theta +t)} - \theta \ln {(\theta)}  = 
    \theta\ln (1+\frac{t}{\theta})\,,
\end{equation}
in accordance with Eq.~(\ref{eq:hoppe_exact}). 

In order to compute the frequency-rank distribution $n(R)$,
let us call $n_i$  the number of balls with  color $i$ in the urn. It follows
the equation:
\begin{equation}
n_i(t+1)=n_i(t)+ \frac{n_i(t)}{\theta +t} \,,
\end{equation}
where the last term is the probability of extracting the
color $i$ at step $t+1$. Taking again the continuous limit: 
\begin{equation}
\frac{dn_i(t)}{  dt } = \frac{n_i}{\theta+t} ~,   ~~~~ n(t_i)=1 \,,
\end{equation}
where the initial conditions are given at the time $t_i$
when the color $i$ first entered in the urn.
By solving the differential equation with the given initial conditions,
we obtain the result
\begin{equation}
n_i (t)= \frac{\theta +t}{\theta +t_i} \simeq \frac{t}{\theta +t_i}  \,,
\end{equation}
where the last approximation holds for $t \gg \theta$.
To obtain the probability distribution of the color
frequencies, we can now consider
\begin{eqnarray}
P(n_i<n) &=&P(t_i >\frac{t}{n} -\theta) = 1 - P(t_i <\frac{t}{n}
-\theta) \simeq 1- \frac{D(\frac{t}{n}-\theta)}{D(t)} = \nonumber \\
&=&  1-
\frac{\theta \ln{\frac{t}{n}} -\theta \ln{\theta}}{ \theta
  \ln{t+\theta} - \theta \ln{\theta} } \simeq 1-
\frac{\theta \ln{\frac{t}{n}} -\theta \ln{\theta}}{ \theta
  \ln{t} - \theta \ln{\theta} } =  \frac{\ln {n}}{ \ln {t}-\ln{\theta}}\,,
\end{eqnarray}
and thus
\begin{equation}
p(n)= \frac{\partial P(n_i<n)}{\partial n} \propto \frac{1}{n} \,.
\end{equation}
Finally, the rank $R$ of a color with $n$ occurrences can
be obtained by considering
\begin{equation}\label{eq:r_exp}
R -1 \simeq k \int_{n}^{n_{\max}} p(n') dn'  = k A \ln{\frac{n_{\max}}{n}}   \,,
\end{equation}
 where $k$ is the number of different colors in the urn and
$A$ is the normalization constant of the probability distribution  $p(n)$:
\begin{equation}\label{eq:A}
A^{-1} = \int_{n_{\min}}^{n_{\max}} \frac{1}{n'} dn'  =\ln
\frac{n_{\max}}{n_{\min}} = \ln{n_{\max}}\,,
\end{equation}
were we set $n_{\min}=1$. 
By inverting then  Eq.~(\ref{eq:r_exp}) we obtain
\begin{equation}\label{eq:nR_hoppe}
n(R) =n_{\max} \exp{\left(-\frac{R-1}{k}\ln {n_{\max}}\right)} \,,
\end{equation}
with $k$ given by Eq.~(\ref{eq:Dt_hoppe}). 
In order to estimate $n_{\max}$, let us write the normalization condition
\begin{equation}\label{eq:intnR}
 \int_{1}^{k} n(R') dR'  =t \,.
\end{equation}
By inserting Eq.~(\ref{eq:nR_hoppe})
in Eq.~(\ref{eq:intnR}),  and noting that
\begin{equation}
n(k) = 1 = n_{\max} \exp{\left(-\frac{k-1}{k}\ln {n_{\max}}\right)}  \,,
\end{equation}
one obtains the relation:
\begin{equation}
k= t \frac{\ln {n_{\max}}}{n_{\max} -1} \simeq  t \frac{\ln
  {n_{\max}}}{n_{\max}} \,, 
\end{equation}
which, by inserting Eq.~(\ref{eq:Dt_hoppe}) for
$k$, gives the result $ n_{\max} \simeq t/\theta$.  We thus obtain the
estimate:
\begin{equation}\label{eq:nR_hoppe_final}
n(R) \simeq \frac{t}{\theta} \exp{\left(-\frac{R-1}{\theta} \right)} \,.
\end{equation}
Note that the frequency-rank distribution $n(R)$ was equivalently indicated as $f(R)$ in previous sections.

\begin{figure*}[t]
\centering
\includegraphics[width=0.47\textwidth]{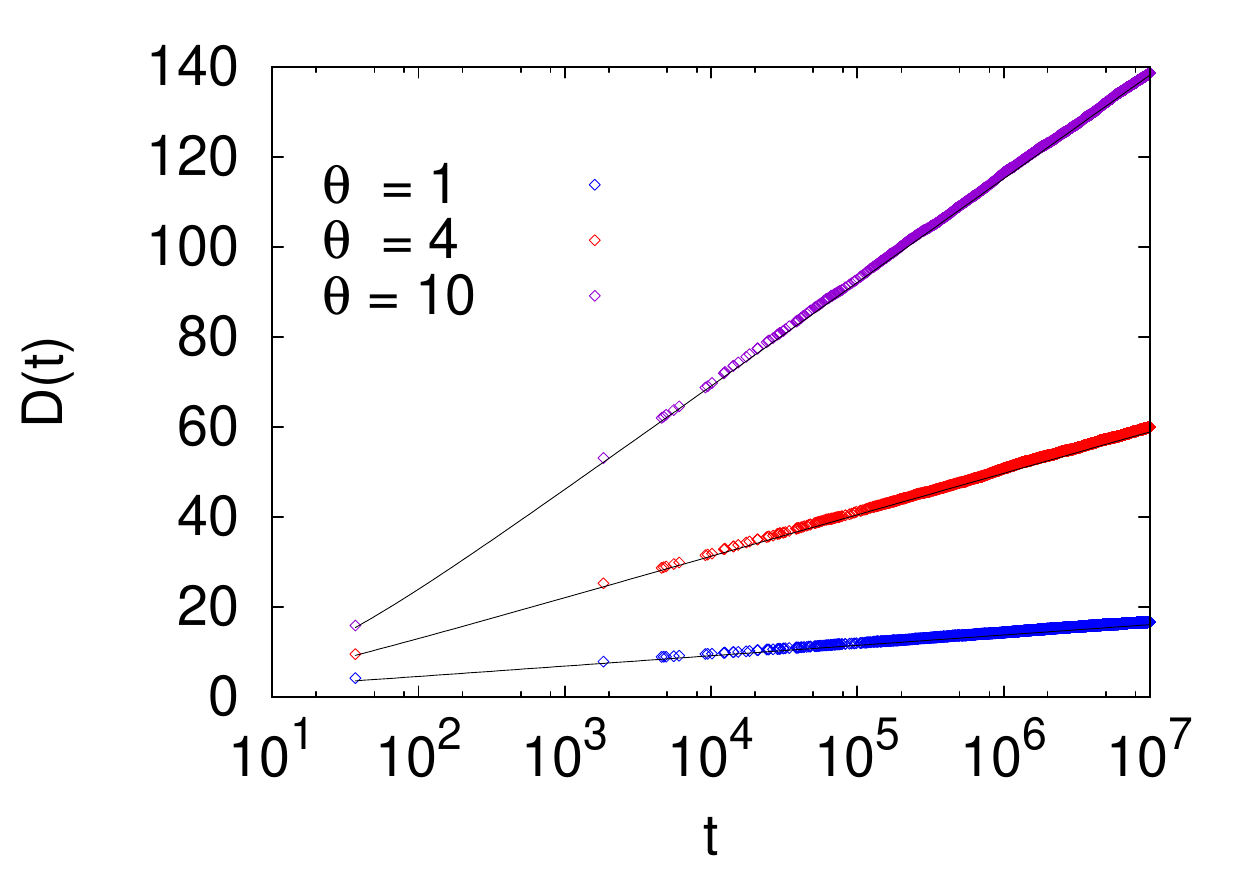}%
\includegraphics[width=0.47\textwidth]{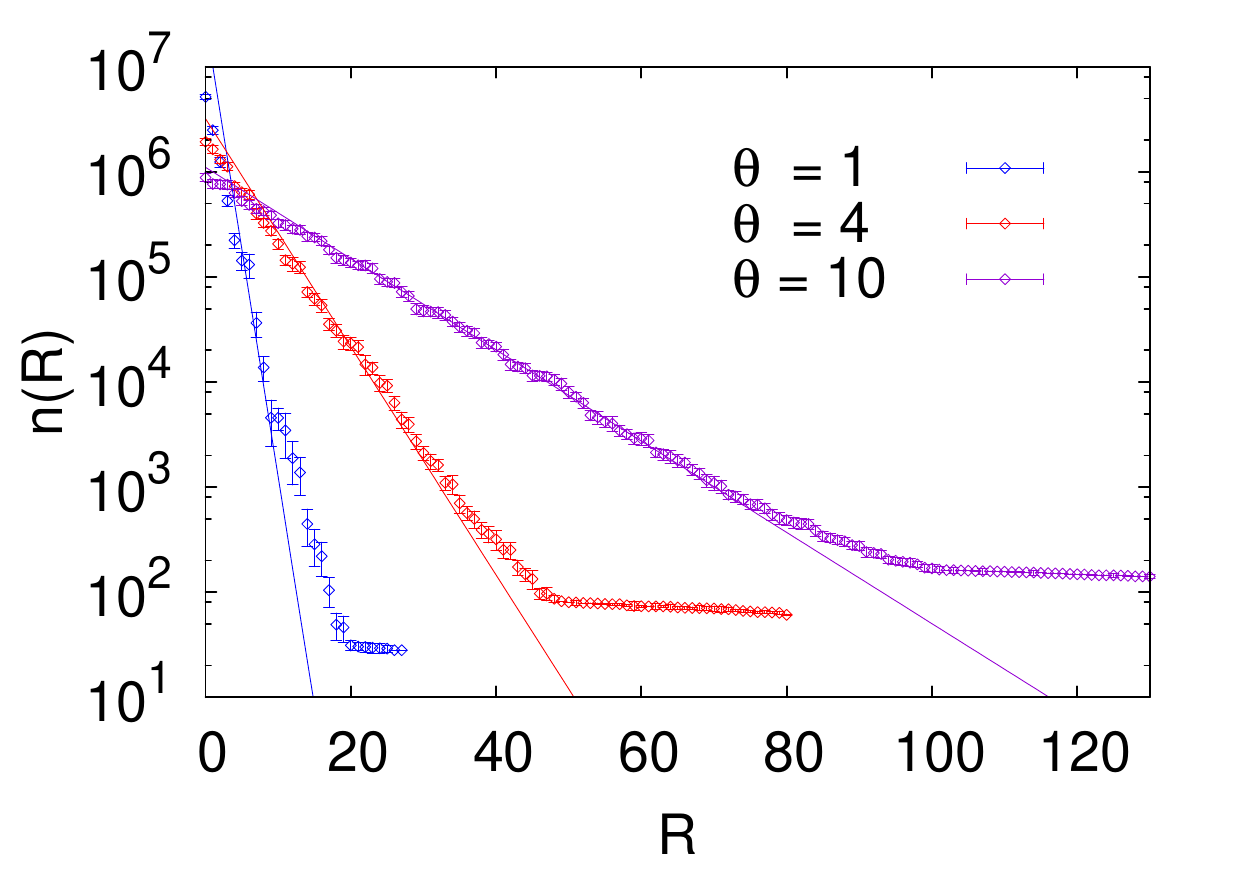}
\caption{ {\bf Zipf's and Heap's laws in the Hoppe urn model.}
 Left:
  Heap's law in the Hoppe urn model for different values of the black
  ball mass $\theta$. Curves correspond to averages
  over $100$ processes. Straight lines correspond to
  Eq.~(\ref{eq:Dt_hoppe}), for each value of $\theta$.
Right:
  Zipf's law in the Hoppe urn model for different values of the black
  ball mass $\theta$. Curves correspond to averages
  over $100$ processes (standard errors are reported). Straight lines correspond to
  Eq.~(\ref{eq:nR_hoppe_final}), for each value of $\theta$, and for the
  length of the process $t=10^7$. 
\label{fig:Heap_hoppe}
  }
\end{figure*}

In Fig.~\ref{fig:Heap_hoppe}  we show results of the Zipf's and Heap's laws,
together with their analytic approximations.
Again, this model predicts a novelty rate of appearance that is much slower than what is found in many real systems, since the number of
different colors $D(t)$ in the urn grows only logarithmically with the number of
extractions $t$.

\section{Urn model with triggering}
\label{sec:vittorio_polya_urns}
In this section we generalize the urn models seen in the preceding section to propose a modeling scheme that incorporates the notion of the adjacent possible so that one novelty can trigger further novelties. Our approach thus builds on that of Hoppe~\cite{Hoppe_1984} and other researchers (see Refs.~\cite{Kotz_1996, Alexander_2012} and references therein), who introduced novelties within the framework of Polya's urn but did not posit that they could trigger subsequent novelties.  Hoppe's model was motivated by the biological phenomenon of neutral evolution, with novel alleles represented as an open-ended set of colors arising via mutation from a single fixed color. This variant of Polya's urn implies a logarithmic, rather than power-law, form for the growth of new colors in the urn, and hence does not account for Heaps' law. Hoppe's urn scheme is non-cooperative in the sense that no conditional appearance of new colors is taken into account; in particular, one novelty does nothing to facilitate another. In 
contrast, the cooperative triggering of novelties is essential to our model.

\subsection{Model definition}
\begin{figure*}[t]
\centering
\includegraphics[width=0.95\textwidth]{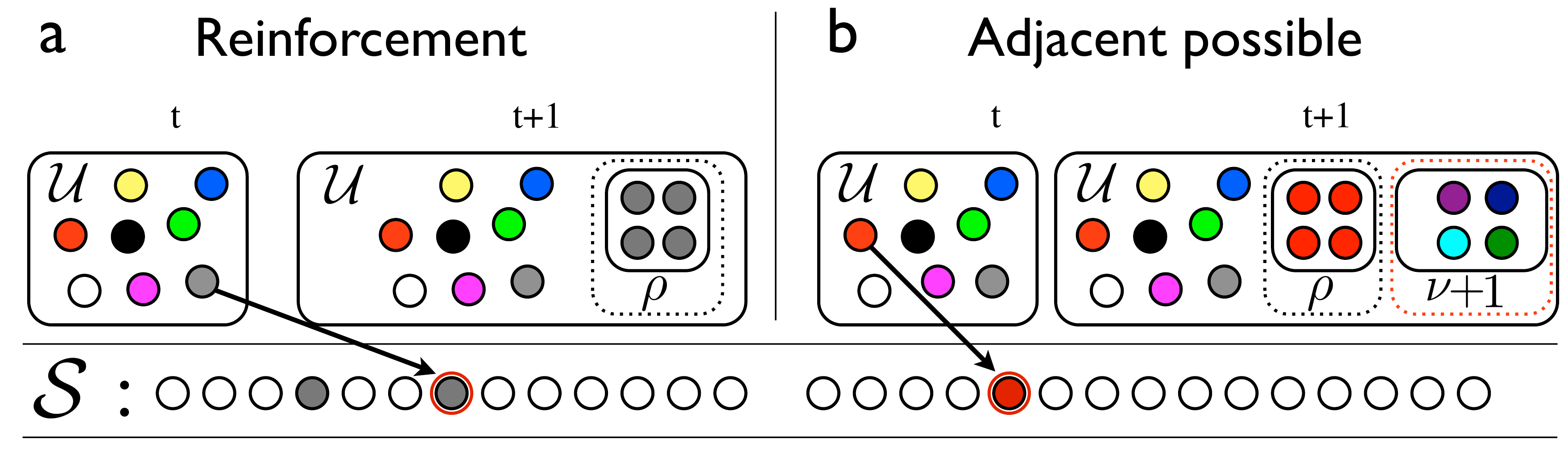}
\caption{ {\bf Models.} Simple urn model with triggering. {\bf (a)} Generic reinforcement step of the evolution. An element (the gray ball) that had previously been drawn from the urn {\em U} is drawn again. In this case one adds this element to {\em S} (depicted at the center of the figure) and, at the same time, puts $\rho$ additional gray balls into {\em U}. {\bf (b)} Generic adjacent possible step of the evolution. Here, upon drawing a new ball (red) from {\em U}, $\nu+1$ brand new balls are added to {\em U} along with the $\rho$ red balls of the reinforcement step that takes place at each time step.}
\label{fig:rules}
\end{figure*}

Consider an urn {\em U} initially containing $N_0$ distinct elements, represented by balls of different colors (Fig.~\ref{fig:rules}). These elements represent songs we have listened to, web pages we have visited, inventions, ideas, or any other human experiences or products of human creativity. A series of inventions is idealized in this framework as a sequence {\em S} of elements generated through successive extractions from the urn. Just as the adjacent possible expands when something novel occurs, the contents of the urn itself are assumed to enlarge whenever a novel (never extracted before) element is withdrawn. Mathematically we consider an ordered sequence {\em S}, constructed by picking elements (or balls) from a reservoir (or urn), {\em U},  initially containing  $N_0$ distinct elements.  Both the reservoir and the sequence increase their size according to the following procedure. At each time step: 
\begin{enumerate}
\item[(i)] an element is randomly extracted from {\em U} with uniform probability and added to  {\em S};
\item[(ii)] the extracted element is put back into {\em U} together with $\rho$ copies of it;
\item[(iii)] if the extracted element has never been used before in {\em S} (it is a new element in this respect), then $\nu+1$ different brand new distinct elements are added to {\em U}.
\end{enumerate}
Note that the number of elements $N$ of {\em S}, i.e.\ the length $|\mathcal{S}|$ of the sequence, equals the number of times $t$ we repeated the above procedure.  If we let $D$ denote the number of distinct elements that appear in {\em S}, then the total number of elements in the reservoir after $t$ steps is $|\mathcal{U}|_t=N_0+(\nu+1) D+\rho t$.

In parallel with the previous one we consider a slightly different variant of the model, in which the reinforcement does not act when an element is chosen for the first time.  Hence, point (ii) of the previous rules will be changed into:
\begin{enumerate}
\item[(ii.a)] 
	the extracted element is put back in {\em U} together with $\rho$ copies of it \emph{only if it is  not new in the sequence}.
\end{enumerate}

\subsection{Computation of the asymptotic Heaps' and Zipf's laws}

We discuss here the asymptotic behavior of both the number of distinct elements $D(t)$ appearing in the sequence  and the
frequency-rank distribution $f(R)$ of the elements in the sequence {\em S}.  We will show that both versions of the urn model above predict a Heaps' law for $D(t)$ and a frequency-rank distribution $f(R)$ with a fat-tail behavior. Our calculations yield simple formulas for the Heaps' law exponent and the exponent of the asymptotic power-law behavior of the frequency-rank distribution  in terms of  the model parameters $\rho$ and $\nu$. 

Strictly speaking, Zipf's law requires an inverse proportionality between the frequency and rank of the considered quantities~\cite{zipf_1949}. In the following, however, we shall always refer instead to a generalized version of Zipf's law, in which the dependence of the frequency on the rank is  power-law-like in the tail of the distribution, i.e.\ at large ranks. 

\paragraph{\underline{Heaps' law}}
In the first version of the model, the time dependence  of the number  $D$  of different elements in the sequence {\em S} obeys the following differential equation: 

\begin{equation}
  \frac{d D}{dt}= \frac{U_D(t)}{U(t)}=\frac{N_0 + \nu D}{N_0+(\nu+1) D+\rho t},\label{eq:D_complete}
\end{equation}
 where $U_D(t)$ is the number of  elements in the reservoir at time $t$ that have not yet appeared in {\em S}, and $U(t)=|\mathcal{U}|_t$ is the total number of elements in the reservoir at time $t$. The term $\nu D$ in the numerator of the rightmost expression comes from the fact that each time a new element is introduced in the sequence,  $U_D(t)$ is increased by  $\nu$  elements (since $\nu+1$ brand new elements   are added to {\em U}, while the chosen element is no longer new). Due to the inherently discrete character of $D$ and $t$, Eq.~(\ref{eq:D_complete}) is valid asymptotically for large values of $D$ and $t$.

In the second version of the model, Eq.~(\ref{eq:D_complete}) has to be modified by replacing the denominator with
\[
	U(t) = N_0+(\nu+1) D+\rho (t-D)= N_0+(\nu+1-\rho) D+\rho t.
\]

To analyze both versions of the model simultaneously, it is convenient to define a parameter $a \equiv \nu+1$ for the first version and $a \equiv \nu+1-\rho$ for the second version.

In order to obtain an analytically solvable equation, and  since we are interested in the behaviour at large times $t \gg N_0$, we approximate equation (\ref{eq:D_complete}) by 
\begin{equation}
\frac{d D}{dt}= \frac{\nu D}{ a D+ \rho t}.\label{eq:D}
\end{equation}
%
 This equation can be solved by defining $z=\frac{D}{t}$, obtaining:
 
 \begin{equation}
   z^\prime t + z= \frac{\nu z}{a z+\rho} ,\label{eq:zprime}
 \end{equation}
 
 \noindent and thus:
 \begin{equation}
 \int \frac{a z +\rho}{z(\nu-\rho -a z)}dz = \int \frac{dt}{t},\label{eq:z}
 \end{equation}

 \noindent Solving the integral we obtain:
 
 \begin{equation}
   \frac{\rho}{\nu-\rho}\log{z}  - \frac{\nu}{\nu-\rho}\log{\left( z a+\rho-\nu\right)} =\log{t} \nonumber
 \end{equation}
 
 \noindent and after some algebra:

 \begin{equation}
 \frac{z^{\frac{\rho}{\nu}}}{z a+\rho-\nu}=t^{\frac{\nu-\rho}{\nu}} 
 \end{equation}
 
 \noindent By substituting $z=\frac{D}{t}$ and again after some algebra we obtain:
 
 \begin{equation}
 D^{\frac{\rho}{\nu}}- a D = (\rho-\nu) t,
 \label{eq:Dfinal}
 \end{equation}
%
from which we can derive the asymptotic behaviour of $D(t)$ for large $t$:
\begin{enumerate}
\item $\rho > \nu$: $D \sim (\rho-\nu)^{\frac{\nu}{\rho}} t^{\frac{\nu}{\rho}}$;\\
\item $\rho < \nu$: $D \sim \frac{\nu-\rho}{a}t$;\\
\item $\rho = \nu$: $D \log D \sim \frac{\nu}{a} t \rightarrow D \sim \frac{\nu}{a} \frac{t}{\log t}$,\\
\end{enumerate}

\noindent where the last estimate cannot be deduced directly from equation~(\ref{eq:Dfinal}), but is deduced by substituting $\nu=\rho$ directly in equation~(\ref{eq:zprime}).

In the case $\rho<\nu$ we recover the results of the well-known Simon's Model~\cite{simon-mandel-dispute}, originally proposed in the context of linguistics and described in section~\ref{sec:vito_simon}. The Simon's model leads to a Zipf's law with an exponent $-(1-p)$ compatible with a linear growth in time of the number of different words. In the framework of the present {\em urn model with triggering} we recover the same Zipf's exponents as well as the linear growth of $D(t)$ if $p=1-\frac{\rho}{\nu}$, with $\rho<\nu$\footnote{We note that if $\nu \gg 1$ when $a=\nu+1$ (first version of the model) or $\nu \gg \rho$ and $\nu \gg 1$ when $a=\nu+1-\rho$ (second version of the model) our model also reproduces the same prefactor of the linear growth of $D(t)$ as in Simon's model. This is evident by setting $a=\nu$ in Eq.~(\ref{eq:D}).}.

For completeness, we note that both versions of the model can be regarded as the coarse-grained  equivalent of a two-color asymmetric Polya urn model~\cite{mahmoud_polya}. In particular, within that finer framework the substitution matrices (denoted $M_1$ for the first version of the model and $M_2$ for the second) would be:
\[ 
	M_1=\left( \begin{array}{cc}
		\rho & 0 \\
		1+\rho & \nu \end{array} \right)
~~\mbox{and}~~ 
	M_2=\left( \begin{array}{cc}
		\rho & 0 \\
		1 & \nu \end{array} \right).
\] 
In this interpretation, the elements that have already appeared in {\em S} are represented by balls of one color, while those that have not appeared yet correspond to balls of the other color.

\paragraph{\underline{Zipf's law}}

Making the same approximations as above, the continuous dynamical equation for the number of occurrences $n_i$ of an element $i$ in the sequence {\em S} can be written as
\begin{equation}
	\frac{d n_i}{d t} = \frac{n_i \rho  +1}{N_0+aD+\rho t}\cdot
\label{eq:dni_dt}
\end{equation}
Two cases can be distinguished: 
\begin{enumerate}
\item $\nu\leq \rho$, when  
\( \displaystyle \lim_{t  \rightarrow +\infty} D/t =  0\). 
By considering only the leading term for \mbox{$t \rightarrow +\infty$}, one has
\begin{equation}
	\frac{d n_i}{d t} \simeq \frac{n_i}{t}.
\end{equation}
Let $t_i$ denote the time at which the element $i$ occurred for the first time in the sequence. Then the solution for $n_i(t)$ starting from the initial condition  $n_i(t_i)=1$ is given by
%
\begin{equation}
	n_i =\frac{t}{t_i}. \label{eq:ni}
\end{equation}
Now consider the cumulative distribution $P(n_i\leq n)$. From Eq.~(\ref{eq:ni}), we can write $P(n_i\leq n)=P(t_i \geq \frac{t}{n})= 1 -P(t_i <\frac{t}{n})$. This leads to the estimate: 
\begin{equation}
P(t_i<\frac{t}{n})\simeq\frac{D(\frac{t}{n})}{D(t)}= n^{-\frac{\nu}{\rho}}. \label{eq:cumul1}
\end{equation}

\item $\nu>\rho$, when  $D \simeq \frac{\nu-\rho}{a} t$. Again considering $t \gg N_0$, we write:
\begin{equation}
\frac{d n_i}{d t} \simeq \frac{\rho n_i}{(\rho+a \frac{\nu-\rho}{a})t} = \frac{\rho n_i}{\nu
  t},
\end{equation}
which yields the solution
\begin{equation}
n_i = \left(\frac{t}{t_i} \right)^{\frac{\rho}{\nu}} .
\end{equation}
Proceeding as in the previous case, we find 
\( 
	P(n_i\leq n)=P(t_i \geq	 {t}\,{n^{-\frac{\nu}{\rho}}})=1 -P(t_i< {t}\,{n^{-\frac{\nu}{\rho}}})
\), 
and thus 
\begin{equation}
	P(t_i< {t}\,{n^{-\frac{\nu}{\rho}}})\simeq \frac{D({t}\,{n^{-\frac{\nu}{\rho}}})}{D(t)}= n^{-\frac{\nu}{\rho}},\label{eq:cumul2}
\end{equation}
obtaining the same functional expression of the asymptotic power-law behavior of the frequency-rank distribution as in the previous case.
\end{enumerate}
The probability density function of the occurrences of the elements in the sequence is therefore
\(
	P(n)= \frac{\partial P(n_i<n) }{\partial n}
	\sim n^{-\left(1+\frac{\nu}{\rho}\right)}
\), 
which corresponds to a frequency-rank distribution $f(R) \sim R^{-\frac{\rho}{\nu}} $.

\begin{figure*}[t]
\centering
\includegraphics[width=\textwidth]{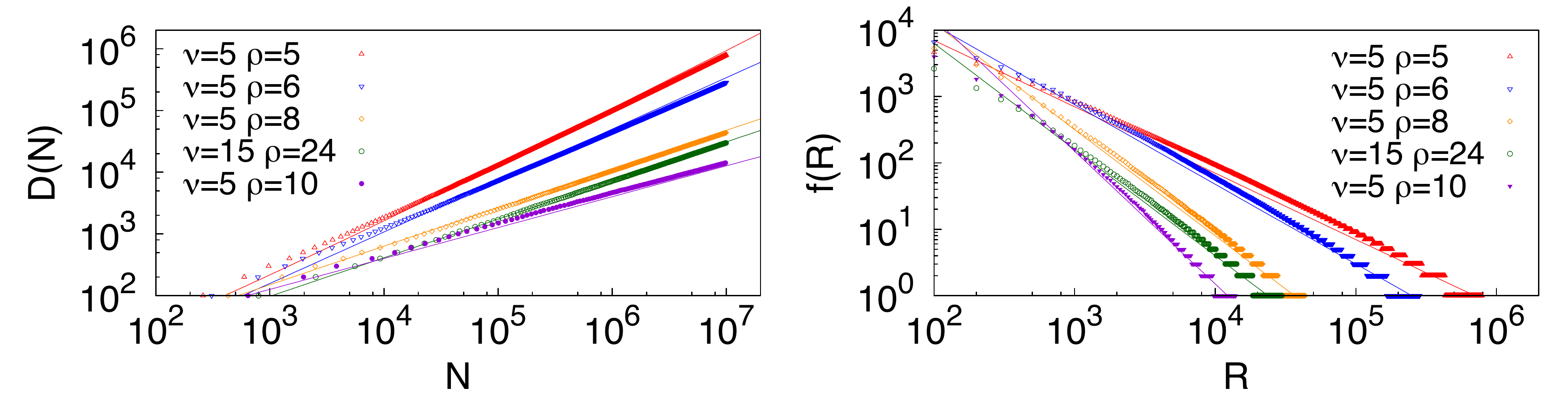}
\caption{{\bf Heaps' law (left) and Zipf's law (right) in the urn model with triggering.}  Straight lines in the Heaps' law plots show functions of the form $f(x)=a x^{\gamma}$ with the exponent $\gamma=\nu/\rho$ as predicted by the analytic results and confirmed in the numerical simulations. Straight lines in the Zipf's law plots show functions of the form $f(x)=a x^{-\alpha}$, where the exponent $\alpha$ is equal to $\gamma^{-1}$.  Note that the frequency-rank plots in real data deviate from a pure power-law behavior and the correspondence between the $\gamma$ and $\alpha$ exponents is valid only asymptotically.}
  \label{fig:model_heaps_zipf}
\end{figure*}

Fig.~\ref{fig:model_heaps_zipf} shows the theoretical predictions  and the numerical simulations for the Heaps' and Zipf laws.  The robustness of the results with respect to fluctuations of the model parameters $\nu$ and $\rho$ was also checked by sampling their values out of several probability distributions.  Note that the estimates in equations~(\ref{eq:cumul1}) and~(\ref{eq:cumul2}) have been derived under the assumption that $t/n \gg1$, i.e.\ in the tail of the frequency-rank distribution. In this respect, it is important to recognize that Zipf's and Heaps' laws are not trivially and automatically related, as is sometimes claimed. We certainly agree that Heaps' law can be derived from Zipf's law by the following random-sampling argument: if one assumes a strict power-law behaviour of the frequency-rank distribution $f(R) \sim R^{-\alpha}$ and constructs a sequence by randomly sampling from this Zipf distribution $f(R)$, one recovers Heaps' law with the functional form $D(t) \sim t^{\gamma}$ with $\gamma =1/\alpha$~\cite{serrano_2009,lu_2010}. 
But the assumption of random sampling is strong and sometimes unrealistic. If one relaxes the hypothesis of random sampling from a power-law distribution, the relationship between Zipf's and Heaps' law becomes far from trivial. In our model, and in work by others~\cite{lu_2010}, the relationship $\gamma = 1/\alpha$ holds only asymptotically, i.e.\ only for large times, with $\alpha$ measured on the tail of the frequency-rank distribution. 

\section{Conclusions}
The processes leading to the emergence of novelties and innovations are mostly unknown. Still, the observation of statistical regularities displayed by the occurrence of new events is key to chart the unknown territories that describe the space of possibilities for societies, biological systems and technology. In this short and far from exhaustive review, we made an attempt to draw a path through the attempts recently made to model, with tools borrowed from the theory of complex systems, the emergence of novelties and innovations. 
To this end we considered a strip of the most renowned models, proposed in almost a century of studies. Some models were historically treated without the explicit aim of modeling the occurrence of novelties and innovations. For example, the model of Simon (see Section~\ref{sec:vito_simon}) was conceived to reproduce the frequency distribution words display in texts and in fact in Simon's model the rate of novelty creation is constant, while in many real systems, including texts, it decreases in time with a power-law behavior. Nevertheless, in light of the results of the new model of Polya's urn with triggering (Section~\ref{sec:vittorio_polya_urns}), Simon's model turns out to be correctly describing a case in which the space of possibilities grows at a fast pace ($\nu>\rho$). By ad-hoc inserting a sub-linear rate of inventions in Simon's model, Zanette and Montemurro~\cite{zanette2005} obtained a satisfactory description of both the frequency distribution of words in texts and the rate of occurrence of new words. Of course the latter is a trivial consequence of having imposed the correct rate \emph{deus ex machina}. Simon's model can be considered a milestone for all models based on the construction of a stream of tokens (e.g., those models involving memory effects, non-linear preferential attachment, etc.) and its limited ability in reproducing real data  has been perfected with the idea of using Polya's urns. Polya's urns are well known to mathematicians who developed a multitude of techniques to cope with them, for instance looking at urns containing balls with a finite number of colors, i.e., with a finite space, and nontrivial transition probabilities. One step forward is that of introducing a simple way to enlarge the space of possibilities. To this end, the Hoppe-Polya model (Section~\ref{sec:fra_hoppe}) already represents a good solution, though the rate of occurrence of innovations is still too low and far from the actually observed values in many systems of interest. Finally, the model of Polya's urn with innovation triggering (Section~\ref{sec:vittorio_polya_urns}), formalizing the notion of adjacent possible envisioned by S.~Kauffman, presents for the first time a satisfactory first-principle based way of reproducing empirical observations. Not only both Heaps' and Zipf's laws of real case situations are reproduced, but also the classical model of Simon is retrieved in the limit of  fast growth of the space. 

In a somewhat humorously self-referring sense, each proposed model has been in the adjacent possible of the models prior to it. But of course this is only an a posteriori consideration. Nobody knows what the adjacent possible space looks like and even conceptually it is not clear which tools one could possibly adopt to chart it. From this perspective we hope that the recent stream of investigations connected to the Polya's urn presented in Section~\ref{sec:vittorio_polya_urns}, by providing the first quantitative characterization of the dynamics of correlated novelties, could be a starting point for a deeper understanding of the different nature of triggering events (timeliness, scales, spreading, individual vs. collective properties) along with the signatures of the adjacent possible at the individual and collective level, its structure and its restructuring under individual innovative events.

\section{Appendix}
\label{sec:fra_zipf_heaps}
\newcommand{\balpha}{\beta}%
\newcommand{\abeta}{\alpha}%
\paragraph{\textbf{Relation between frequency distribution and frequency rank distribution (Zipf's law)}}

Let us consider a sequence of $N$ random variables (or letters, or any kind of item),  
and the frequency at which each particular value enters in the sequence. For many systems (genes in a pan-genome, words in a texts, etc..), the distribution of the number of appearances of the same value of the variable in a sequence is a power law:

\begin{equation}\label{eq:p_f}
p(f) =(\balpha -1) f_{\min}^{\balpha-1} f^{-\balpha} \,,
\end{equation}

\noindent with $\balpha > 1$.

Let us now relate the {\em frequency distribution} to the {\em frequency rank distribution}:
the elements in the sequence are ordered according to their frequency in decreasing order (rank one for the most frequent element), and the frequency is studied as a function of the rank $R$ of each element.

Let us now compute the form of frequency rank curve (also called Zipf's law) corresponding to the frequency distribution in \Eq{eq:p_f}.
The rank $R$ of an element with frequency $f$ is defined as the number of different elements with frequency $\tilde{f} \geq f$:

\begin{equation}\label{eq:rank}
R(f) \simeq  k \int_{f}^{+\infty} p(\tilde{f}) d\tilde{f}   \,,
\end{equation}

\noindent where $k$ is the number of distinct elements in the sequence. 
By substituting \Eq{eq:p_f} into \Eq{eq:rank} we obtain:

\begin{equation}\label{eq:f_rank}
R(f)\propto    f^{1-\balpha}    
\end{equation}

\noindent and inverting the relation:

\begin{equation}
 f(R) \propto R^{-\abeta}  \,\,\,\,\,\,\,\,\,\,\, \text{with} \,\,\,\,\,\,\,\,\,\,\, \abeta = \frac{1}{\balpha -1} \, .
\end{equation}

\noindent In order to obtain the correct expression for $f(R)$, we now have to consider the normalization:

\begin{equation}
\int_{1}^{R_{\max}} f(\tilde{R}) d  \tilde{R} = 1 \,.
\end{equation}

\noindent Let us now distinguish the cases:

\begin{description}

\item[] $\abeta \neq 1$ ($\balpha \neq 2$):

\begin{equation} \label{eq:zipf}
 f(R) =  \frac {1-\abeta}{R_{\max}^{1-\abeta} -1}  R^{-\abeta} \,.
\end{equation}

\item[] $\abeta = 1$  ($\balpha=2$):

\begin{equation}
 f(R) = \frac {1} {\ln{R_{\max}}} R^{-1} \,.
\end{equation}

\end{description}

\noindent Let us now note that when $\abeta >1$, we can neglect the term $R_{\max}^{1-\abeta}$ in \Eq{eq:zipf}, and when $\abeta <1$, we can write $R_{\max}^{1-\abeta} -1 \simeq R_{\max}^{1-\abeta} $ .

Summarizing:

\begin{eqnarray}
\abeta >1 \,\,\, (\balpha<2): && \,\,\,\,\,\,
f(R) \simeq  (\abeta -1)  R^{-\abeta} \,. \label{eqn1:zipf}\\
&& \nonumber \\
\abeta = 1  \,\,\,  (\balpha=2): &&\,\,\,\,\,\,
 f(R) \simeq \frac {R^{-1}} {\ln{R_{\max}}} \,. \label{eqn2:zipf}\\
&& \nonumber \\
\abeta<1\,\,\,  (\balpha>2): && \,\,\,\,\,\, 
 f(R) \simeq (1-\abeta) \frac {R^{-\abeta}}{R_{\max}^{1-\abeta} }  \,.\label{eqn3:zipf}
\end{eqnarray}

\paragraph{\textbf{Relation between frequency rank distribution and number of distinct elements (Heaps law)}}

We now want to estimate the number $D$ of distinct elements appearing in the sequence as a function of its length $N$
(Heaps' law). To do that, let us consider the entrance of a new element (never appeared before) in the sequence and let the number of distinct elements in the sequence be $D$  after this entrance, and the length of the sequence $N$. This new element will have maximum rank $R_{\max} =D$, and frequency $f(  R_{\max}) = 1/N$.  
From Eqs.~(\ref{eqn1:zipf}), (\ref{eqn2:zipf}), (\ref{eqn3:zipf}) we thus obtain:

\begin{eqnarray}
\abeta >1 \,\,\, (\balpha<2): && \,\,\,\,\,\,
f(D) \simeq  (\abeta -1)  D^{-\abeta}  = \frac{1}{N}\,. \\
&& \nonumber \\
\abeta = 1  \,\,\,  (\balpha=2): &&\,\,\,\,\,\,
 f(D) \simeq \frac {1} {D \ln{D}} = \frac{1}{N}\,. \\
&& \nonumber \\
\abeta<1\,\,\,  (\balpha>2): && \,\,\,\,\,\, 
 f(D) \simeq \frac {1-\abeta}{D^{1-\abeta} -1}  D^{-\abeta} = \frac{1}{N} \,.
\end{eqnarray}

\noindent Inverting these relations we finally find:

\begin{eqnarray}
\abeta >1 \,\,\, (\balpha<2): && \,\,\,\,\,\,
D \simeq N^{\gamma} \,\,\,\,\,\,\,\,\,\,\, \text{with} \,\,\,\,\,\,\,\,\,\,\,  \gamma= \frac{1}{\abeta}\,. \label{eqn1:heaps}\\
&& \nonumber \\
\abeta = 1  \,\,\,  (\balpha=2): &&\,\,\,\,\,\,
 D \simeq \frac{N}{\ln{N}}\,. \label{eqn2:heaps}\\
&& \nonumber \\
\abeta<1\,\,\,  (\balpha>2): && \,\,\,\,\,\, 
 D \simeq N \,.\label{eqn3:heaps}
\end{eqnarray}
\paragraph{\textbf{Some remarks about text generation}}
The production of written text in a given idiom is a peculiar process.
Although texts are generally taken as a paradigmatic example of a collection of tokens (words) that obey both Zipf's and Heaps' laws and many modelling efforts have been undertaken to explain this feature, writers very seldom create brand new words in that idiom.
Rather, the process of text production is well approximated by the sampling of the existing Zipf's law of that idiom at the specific time of its writing.
Authors, then, choose words according to their frequency and add the necessary short range correlations needed to compose meaningful sentences and long range correlations to follow the plot line.

The generation of the Zipf's law of words in a given idiom is therefore a different process with respect to book writing and the scientific data analysis and modelling efforts have to be steered toward the comprehension of this global shared law.
Notably, some work has been already done based on \emph{Google ngrams} and it was shown that the Zipf's distribution of English varies from decade to decade~\cite{altmann2013} but in a way that its functional form stays constant.
The functional form is that of a double slope in logarithmic scale (see Fig.~\ref{fig:heaps_zipf}.d for the Zipf's law inferred from the Gutenberg corpus) with the Heaps' law connected to the tail of the distribution, i.e., to the appearance of less frequent words, as it should be. 
A satisfactory model accounting for the frequency-rank distribution of words in a language has to include different ingredients, therefore, from the mechanism of creation of neologisms, to the ageing and disappearance of terms and revival of others.
Above all, the mechanism has to highlight the collaborative and shared character of idioms.



\end{document}